\documentclass[]{JHEP}

\makeatletter

\@ifundefined{lesssim}{\def\lesssim{\mathrel{\mathpalette\vereq<}}}{}
\@ifundefined{gtrsim}{}{}

\def\vereq#1#2{\lower3pt\vbox{\baselineskip1.5pt \lineskip1.5pt

\ialign{$\m@th#1\hfill##\hfil$\crcr#2\crcr\sim\crcr}}}

\makeatother

\title{Oscillating scalar-field dark matter in supergravity}
\author{Osamu Seto, Kazunori Kohri and Takashi Nakamura \\ 
Yukawa Institute for Theoretical Physics, Kyoto
University, Kyoto 606-8502, Japan }

\abstract{
We show that an oscillating scalar field in supergravity of mass of
the order of $\sim$ TeV with a nonzero vacuum expectation value ($\sim
10^{10}$ GeV) can be a candidate of cold dark matter
(CDM). To avoid the gravitino problem, we need a low reheating
temperature after the primordial inflation. Then, the energy density
of the oscillating scalar field satisfies all the requirements for CDM
at present in the universe.
}


\preprint{YITP-01-65}

\newcommand{\gev}{ {\rm GeV} }
\newcommand{\tev}{ {\rm TeV} }
\newcommand{\order}{{\cal O}}

\begin{document}

\section{Introduction}
It is widely believed that a significant fraction of energy density in
the universe is in the form of cold dark matter (CDM). Recently it was
reported that the contribution of CDM to density parameter is
approximately $\Omega \simeq 0.3$ (for a review, see
Ref.~\cite{Schindler}). It is one of the most important problems in
cosmology and particle physics to clarify the nature and the origin of
dark matter.

In this paper, we consider stable scalar fields with an electroweak
scale mass $\sim \order(\tev)$ and a large vacuum expectation value
(VEV) $\sim \order(10^{10}) \ \gev$. These scalar fields naturally
appear in supergravity. It is interesting that such a scalar filed can
have a net oscillation energy after inflation because of the coupling
with the inflaton field through K\"{a}hler potential and an additional
SUSY breaking effect during the inflation~\cite{Dine:1984ys,
Bertolami:1987xb}. On the other hand, however, whenever we consider a
model based on supergravity within the framework of inflationary
cosmology, we are faced with some sticky problems. In particular,
``gravitino problem'' would be one of the severest problems in
cosmology~\cite{Pagels:1982ke,Weinberg:1982zq}. To avoid the gravitino
problem, i.e., to restrain the production of gravitinos and the
photodissociation of light elements due to their late-time decays, we
need a low reheating temperature after the primordial
inflation~\cite{Holtmann:1999gd}. In this situation, we show that the
energy density of the oscillation of the scalar field satisfies the
requirements for CDM at present in the universe.

\section{Model}
In supergravity, we know there exist a lot of scalar fields with an
almost flat potential. These scalar fields are expected to acquire
masses of the order of the electroweak scale from a supersymmetry
breaking effect and might have a nonzero VEV.

We consider a chiral superfield $\Phi$ which is the gauge singlet and
contains a scalar field $\phi$. We assume that it has a minimal
K\"{a}hler potential $K = |\Phi|^2M_G^{-2}$ and a superpotential,
\begin{equation}
W = \frac{\lambda}{n+3}\frac{\Phi^{n+3}}{M_G^n} + C,
\end{equation}
where $M_G = M_{Pl}/\sqrt{8\pi} \simeq 2.4 \times 10^{18} \gev$ is the
reduced Planck mass, $\lambda \sim \mathcal{O}($$1$$)$, and $C$ is a
constant.
The superpotential and the K\"{a}hler potential give a scalar
potential,
\begin{equation}
    \label{eq:pot0}
    V(\phi) = -3\frac{C^2}{M_G^2}-2\frac{C^2}{M_G^4}|\phi|^2
    +\frac{n}{n+3}\frac{\lambda
    C}{M_G^2}\frac{\phi^{n+3}+\phi^{*n+3}}{M_G^n}+
    \lambda^2\frac{|\phi|^{2n+4}}{M_{Pl}^{2n}} ,
\end{equation}
at low energy limit($\phi \ll M_G$). In addition, the scalar field
acquires a soft mass ($\sim 1$ TeV) through a SUSY breaking
effect. When we assume that the cosmological constant vanishes at the
VEV, it is natural to take $C \simeq m_{3/2}M_G^2$, where $m_{3/2}$ is
the gravitino mass ($\sim 1$ TeV). Then, we obtain the following low
energy effective potential,
\begin{equation}
    V(\phi) = V_0 - m_0^2 |\phi|^2 +\frac{n}{n+3}\frac{\lambda
    C}{M_G^2}\frac{\phi^{n+3}+\phi^{*n+3}}{M_G^n}+
    \lambda^2\frac{|\phi|^{2n+4}}{M_G^{2n}},
\end{equation}
where the second term means the negative mass term ($m_0^2 \sim
C^2/M_G^4 \sim 1\textrm{TeV}^2$) which comes from the negative mass
squared in Eq.~(\ref{eq:pot0}) and the soft mass.  Notice that this
potential does not have a term $\kappa \phi^4$ with $\kappa \sim
\order(1)$. One finds that the VEV of the scalar field $M \equiv
\langle\phi\rangle$ is given by
\begin{equation}
 \label{eq:vev}
 \frac{M^{n+1}}{M_G^n} =
 \frac{1}{2(n+2)\lambda}\left[\frac{nC}{M_G^2}+
   \sqrt{\left(\frac{nC}{M_G^2}\right)^2+4(n+2)m_0^2}\right],
\end{equation}
and the vacuum energy at $\phi = 0$ becomes
\begin{equation}
    V_0 =\frac{n+1}{n+2}M^2\left[m_0^2+\frac{nC}{2(n+2)(n+3)M_G^2}
      \left(\frac{nC}{M_G^2}+
        \sqrt{\left(\frac{nC}{M_G^2}\right)^2+4(n+2)m_0^2}\right)\right].
\end{equation}
From Eq.~(\ref{eq:vev}), we see that $M \sim \sqrt{m_0 M_G} \sim
\order(10^{10})$ GeV in the case of $n = 1$.

After the primordial inflation, the inflaton field oscillates around
its minimum and dominates the energy density in the universe until the
reheating time $t \sim \Gamma_I^{-1}$, where $\Gamma_I$ is the decay
rate of the inflaton field. While the Hubble expansion rate is large
$H \gg m_0$, the scalar field $\phi$ would be trapped dynamically at
the origin by an additional SUSY breaking effect which is explained in
the following reason~\cite{Dine:1984ys, Bertolami:1987xb}.
In supergravity, the scalar potential of the inflaton field $I$ and the
scalar field $\phi$ is expressed by
\begin{equation}
V(\phi,I) = e^{G(\phi,I)}\left[ G_i(G^i_j)^{-1}G^j -3 \right],
\end{equation}
where $G = K + \ln|W|^2$ and $G_i = \partial G/\partial \phi^i, G^j =
\partial G/\partial\phi_j^*$ and $G_i^j = \partial^2
G/\partial\phi^i\partial\phi_j^*$. During the oscillation of the
inflaton field, the scaler potential is related to the Hubble
expansion rate as $V(\phi,I) \simeq \rho_I \simeq 3M_G^2H^2$ through
the Friedmann equation, because the energy density in the universe is
dominated by the inflaton field. Then, the scalar potential is
modified as
\begin{equation}
V(\phi) \simeq V_0 + 3H^2|\phi|^2,
\end{equation}
and $\phi$ would be trapped at $\phi=0$.

On the other hand, when the Hubble expansion rate becomes smaller than
the mass of the scalar field, i.e., $H \lesssim m_0$, the additional
SUSY breaking effect disappears and the scalar field begins to roll
down to its VEV while the oscillating inflaton field still
dominates in the energy density in the universe  because of
the low reheating temperature in order to avoid the
gravitino problem.

The scalar potential is expanded around the VEV as
\begin{equation}
    \label{eq:expand}
V(x) \sim V''(M) x^2 + \lambda_3 x^3 + \lambda_4 x^4 + ... ,
\end{equation}
where $x$ expresses the deviation from the VEV, i.e., $x = \phi -
M$. The mass squared in the VEV, $V''(M)$, is of the order of $\sim
\order(\tev^2)$ while $\lambda_3$ and $\lambda_4$, are
of the order of $V''(M)/M$ and $V''(M)/M^2$, respectively. Note that the
initial amplitude of the oscillation is approximately
$\order(M)$. Then, we see that the oscillation induced by the mass
term, i.e., the first term in Eq.~(\ref{eq:expand}), dominates the
oscillating energy because $x \lesssim M$.  Therefore, we find that
both the energy density of $\phi$  and  the inflaton field $I$
decrease as  $ \propto a(t)^{-3}$ where $a(t)$ is the scale factor. Namely the
ratio of the energy density of the inflaton field to that of the scalar
field does not change until the reheating time.
This means that the vacuum energy of the scalar field $V_0$ is transfered
to the oscillation energy of $\phi$.

Hereafter we mainly consider the case of n = 1. The energy density of
the scalar field at the reheating time $(t=t_R)$ is estimated as
\begin{equation}
    \rho_{\phi}(t_R) = \left.\frac{\rho_{\phi}}{\rho_I}\right|_{H=m_0}
    \rho_I(t_R),
\end{equation}
with the energy density of the inflaton field,
\begin{eqnarray}
    \label{eq:rho_I}
    \rho_I(t_R) = \pi^2 g_* T_R^4/30,
\end{eqnarray}
where $T_R$ is reheating temperature and $g_*$ is the degree of
freedom in the thermal bath. When the reheating process finished, the
ratio of the energy density of $\phi$ to the entropy density $s$
is estimated as
\begin{eqnarray}
    \frac{\rho_{\phi}}{s} &=& \frac{V_0T_R}{4M_G^2m_0^2} , \nonumber
    \\ &\simeq& 0.5 \times 10^{-9} \textrm{GeV} \left(\frac{T_R}{10^7
      \textrm{GeV}}\right)
    \left(\frac{M}{10^{10}\textrm{GeV}}\right)^2 \label{cons},
\end{eqnarray}
where $s = 2\pi^2 g_* T_R^3/45$. Both $\rho_{\phi}$ and $s$ decrease
as $a(t)^{-3}$, so that the ratio $\rho_{\phi}/s$ is constant unless
the additional entropy is produced after the reheating. If we adopt
inflation models with a low reheating temperature, e.g., $T_R \lesssim
10^7 \gev$, in order to avoid the gravitino
problem~\cite{Holtmann:1999gd}, the energy density of the dark matter
cannot be larger than the critical density and does not overclose
the universe.

The present value of the energy density of dark matter to entropy
density ratio is given by
\begin{eqnarray}
    \label{eq:dm}
    \frac{\rho_{\rm DM}}{s_0} &=&\frac{\Omega_{\rm DM}\rho_{cr}}{s_0}
    \nonumber \\ &\simeq& 3.6 \times 10^{-9}\Omega_{\rm DM} h^2 \gev,
\end{eqnarray} 
where $\rho_{cr}$ is the present critical density of the universe,
$s_0$ is the present entropy density, $\Omega_{\rm DM}$ is the density
parameter of dark matter ($\sim$ 0.3)~\cite{Schindler}, and $h$ is
the present Hubble parameter normalized as $H_0 = 100 h \textrm{km/sec
Mpc}{}^{-1}$. From Eq. (\ref{eq:dm}) we see that the energy density of
the oscillating scalar field almost coincides with the energy density
of dark matter. Namely the contribution of $\rho_{\phi}$ to
density parameter $\Omega$ is estimated as
\begin{eqnarray}
    \Omega_{\phi} &=& \frac{\rho_{\phi}}{\rho_{cr}} \nonumber \\ 
    &\simeq& 0.28 \times \left(\frac{T_R}{10^7
      \textrm{GeV}}\right)\left(\frac{M}{10^{10}
      \textrm{GeV}}\right)^2\left(\frac{0.7}{h}\right)^2
    \label{eq:omega}.
\end{eqnarray}
From Eq.~(\ref{eq:omega}) we see that $\Omega_{\phi} \simeq 0.3 \ 
(0.03)$ for $T_R \simeq 10^7 \ (10^6)$ GeV.  On the other hand, for
the case of $n = 2$, the VEV of the scalar field is of the order of
$10^{13}$ GeV. Then, we obtain $\Omega_{\phi} \simeq 0.3 \ (0.03)$ for
$T_R \simeq 10 \ (1)$ GeV.

\section{Conclusion}

In this paper, within the framework of inflationary cosmology we have
shown that the stable scalar field with the electroweak scale mass and
the large VEV is now oscillating, and the energy density of the
oscillation significantly contributes to the density parameter
$\Omega$ and satisfies the requirements for CDM at present in the
universe. It is interesting that such a scalar field naturally appears
in supergravity. In addition, it is also fascinating that when we
require the low reheating temperature after the primordial inflation
to avoid the gravitino problem ($T_R \lesssim 10^7 \ \gev$), it
automatically ensures the appropriate energy density for CDM
($\Omega_{\phi} \simeq 0.3$).  The potential we considered has
degenerate minima and may cause a domain wall problem. However the
difficulty can be solved by a few modification, as was shown in
Ref.~\cite{Asaka:1999xd}.

\acknowledgments
We would like to thank J. Yokoyama for informative comments.
This work was partially supported by the Grant-in-Aid for Scientific
Research ( Nos.11640274, 09NP0801) from the Japanese Ministry of
Education, Science, Sports, and Culture.

\end{document}